\documentclass[aps,prc,twoside,nofootinbib,showpacs]{revtex4}
\usepackage{amsmath,amssymb}
\usepackage{graphicx}
\newcommand{\beq}{\begin{equation}}
\newcommand{\eeq}{\end{equation}}
\newcommand{\beqa}{\begin{eqnarray}}
\newcommand{\eeqa}{\end{eqnarray}}
\def\pmainz{$^2$}
\def\ptomsk{$^1$}
\def\pminsk{$^3$}
\def\passiut{$^4$}

\begin{document}

\title{\boldmath The role of pion exchange in $\eta$ meson photoproduction
  on the deuteron}

\author{
 A.~Fix\ptomsk\footnote[1]{Electronic address: fix@tpu.ru},
 H.~Arenh\"ovel\pmainz\footnote[2]{Electronic address: arenhoev@kph.uni-mainz.de},
 M.~Levchuk\pminsk\footnote[3]{Electronic address: levchuk@dragon.bas-net.by},
 and M.~Tammam\passiut\footnote[4]{Electronic address: mmatm@yahoo.com}
 \vspace*{0.1in}
}

\affiliation{ \ptomsk Laboratory of Mathematical Physics, Tomsk Polytechnic University,
634034 Tomsk, Russia}%
\affiliation{ \pmainz Institut f\"ur Kernphysik, Johannes Gutenberg-Universit\"at Mainz,
D-55099 Mainz, Germany}%
\affiliation{ \pminsk Stepanov Institute of Physics, Belarus National Academy of Sciences, 220072 Minsk, Belarus}%
\affiliation{ \passiut Physics Department, Al-Azhar University, Assiut, Egypt}%

\date{\today}

\begin{abstract}
Incoherent $\eta$ photoproduction on the deuteron 
is studied with the main emphasis on the role of final state
interactions. In addition to the previously studied mechanisms of
$NN$ and $\eta N$ rescatterings, the role of an intermediate pion
exchange is considered in detail, where first a pion is photoproduced
on one nucleon and then rescatters into an $\eta$ meson on the
other, the spectator nucleon. It is found, that the role of this pion
mediated contribution is comparable in size to that of $\eta N$ rescattering.
Results for total and semi-inclusive differential cross sections and
associated polarization observables are presented. In particular
polarization observables show a significant sensititvity to final
state interactions. 
\end{abstract}


\maketitle

%%%%%%%%%%%%%%%%%%%%%%%%%%%%%%%%%%%%%%%%%%%%%%%%%%%%%%%%%%%%%%%%%%
%%%%%%%%%%%%%%%%%%%%%%%%%%%%%%%%%%%%%%%%%%%%%%%%%%%%%%%%%%%%%%%%%%
\section{Introduction}

\begin{figure}
\begin{center}
\resizebox{0.95\textwidth}{!}{%
\includegraphics{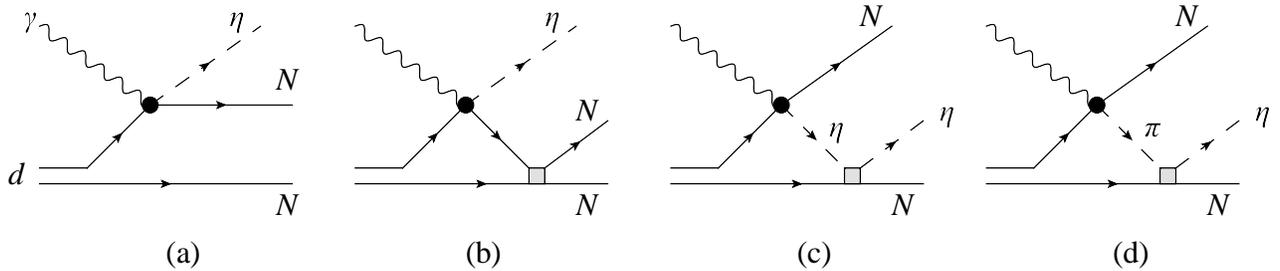}
} \caption{Diagrammatical representation of the amplitude for $\gamma
  d\to\eta np$: (a) Impulse (spectator) approximation (IA), (b) $NN$
  rescattering, (c) $\eta N$ rescattering, and  (d)
pion exchange mechanism ($\pi N$).} \label{fig1}
\end{center}
\end{figure}

Photoproduction of an $\eta$ meson on a deuteron was investigated quite extensively in
Refs.~\cite{Krusche95,FiTR,Sauermann,FiAr97,Sibir,FiAretaNN,FiAr00,FiArNuPh,FiAr04}.
In general, the focus was on the question of how much the influence of final state
interactions (FSI) overshadows the information on the single nucleon
response by studying $\eta$ photoproduction on quasifree
nucleons, in particular, on a neutron. It has been shown that the
interaction effects are rather important in the near-threshold
region but become less significant with increasing photon energy in
the unpolarized total cross section. Thus
at higher energies the impulse
approximation (IA, diagram (a) of Fig.~\ref{fig1}), associated with
the one-nucleon response, provides the dominant part
of the total cross section. However, for the differential cross
section the FSI mechanisms which involve at least two nucleons 
are expected to be of importance in the region of large momentum transfer, primarily at
backward $\eta$ emission angles. Indeed, in this region the energy $\Delta E$ transferred
by the produced meson is essentially lower than the energy transfer associated with
quasifree kinematics, that is $\Delta E \ll {(\Delta p)^2}/{2M_N} $,
where $\Delta p$ is the corresponding momentum transfer and $M_N$ the
nucleon mass. This imbalance results in a
suppression of the single nucleon response in the region of high $\Delta p$ values. In
this situation mechanisms in which two or more nucleons can share the transferred
momentum become increasingly important. As the dominant reaction mode in this kinematic
region one can expect the rescattering of the final two nucleons (diagram (b) of
Fig.~\ref{fig1}). The role of the $\eta N$ rescattering (diagram (c)  of Fig.~\ref{fig1})
was already considered in Refs.~\cite{FiTR,FiAr97,FiAr04} and was shown to be quite
important, too.

If in addition pion degrees of freedom are considered, then $\eta$ photoproduction can proceed
according to the two-step scheme $\gamma N\to\pi N\to\eta N$ (diagram
(d) of Fig.~\ref{fig1}), where the photoproduced
pion is subsequently rescattered into an $\eta$ meson by the spectator
nucleon. To the best of our knowledge, there is no thorough
investigation of this mechanism for this reaction. In general, it was simply ignored, since it was
argued that the corresponding amplitude should be suppressed due to a large momentum of the
intermediate pion. Namely, because the pion has a relatively small mass, its propagation is
associated with a large intermediate momentum and, therefore, should be effective only at
short internucleon distances. For example, the $\pi NN$ channel provides only a small fraction
of the $\eta d$ scattering cross section, which is governed by the long-range structure
of the $\eta NN$ wave function (see Ref.~\cite{FiAretaNN}).

Although these arguments appear reasonable, the insignificance of pion rescattering in
incoherent $\eta$ photoproduction is not obvious since, as already noted above, at higher
energies are kinematic regions where the two-nucleon response becomes essential.
Furthermore, the
cross section for $\gamma N\to\pi N$ is much larger than that for $\gamma N\to\eta N$ so
that the above noted suppression due to a large intermediate momentum can be compensated
by a high yield of pion photoproduction. It is therefore important to study the role of
this pion exchange via a direct inclusion of the corresponding diagram
(d) of Fig.~\ref{fig1} into the reaction amplitude.

Furthermore, the influence of such a pion exchange in kaon photoproduction on a
deuteron was investigated rather detailed in Ref.~\cite{Salam}, where it has been shown
that in some charge channels its inclusion can visibly change even the total cross
section. Namely, in $K^+$ photoproduction the pion mediated process
$\gamma d\to \pi NN \to K\Sigma N$ results in an almost
10-15~$\%$ increase of the total cross section in the maximum and thus seems to be the
most important FSI mechanism in this reaction.

One should note that in some of the previous studies the contribution of such pion
exchange was already partially investigated. However, the
corresponding results suffer from some quite rough approximations. For
example, in
Ref.~\cite{FiArNuPh}, where a three-body $\eta NN-\pi NN$ approach was
used, only the resonance $S_{11}(1535)$ in the pion photoproduction
operator was included. This can lead to a significant underestimation
of the role of the pion exchange, since even in the region of the $S_{11}(1535)$
this resonance plays a very moderate role in pion
photoproduction. Furthermore, in an earlier paper~\cite{FiTR},
a simple approximation with respect
to the choice of energy and momentum of the initial
nucleon in the deuteron was adopted (see the formalism in
Ref.~\cite{FiTR}). Such a treatment was partially justified since the
$\pi$ exchange as well as $\eta$ rescattering are indeed rather
insignificant in the total cross section, to which this paper was
devoted, so that already such a rough estimation of their role was
sufficient to draw some conclusions about the reaction
dynamics. Therefore, in view of these crude approximations for the pion
exchange mechanism in previous studies, it appears timely to bring the
theory to an appropriate quantitative level.

In the next section we briefly describe the most important dynamical
ingredients of the reaction matrix, in particular, those aspects concerning the
pion rescattering mechanism. Sect.~\ref{observables} reviews briefly the general form
of the observables in terms of the $T$-matrix.
In Sect.~\ref{Results} we present our results for the
unpolarized total cross section and associated polarization
asymmetries, the semi-inclusive differential cross section and some selected
corresponding polarization observables, which are especially sensitive to FSI
effects. Some conclusions and  an outlook are given in
Sect.~\ref{Conclusion}.

%%%%%%%%%%%%%%%%%%%%%%%%%%%%%%%%%%%%%%%%%%%%%%%%%%%%%%%%%%%%%%%%%
%%%%%%%%%%%%%%%%%%%%%%%%%%%%%%%%%%%%%%%%%%%%%%%%%%%%%%%%%%%%%%%%%

\section{The $T$-Matrix}\label{tmatrix}

We consider the following reaction
\begin{equation}
\gamma(E_\gamma,\vec{k}\,)+d(E_d,\vec{p}_d\,)\rightarrow
\eta(\omega_\eta,\vec{q}_\eta\,)+N(E_1,\vec{p}_1) + N(E_2,\vec{p}_2)\,,
\end{equation}
where the four-momenta of the participating particles are given in
parenthesis. As reference frame we choose the overall
center-of-mass (c.m.) system with $z$-axis along the incoming
photon momentum and the $x$-axis along the direction of maximal
linear photon polarization.
Total and differential cross sections and associated polarization
observables as well are determined by the $T$-Matrix element
\beq\label{Tsms}
T_{s m_s t \lambda m_d}= -^{(-)}\langle \vec p_1\,\vec p_2\,s
m_s,t,\,\vec q\,| \vec\varepsilon_\lambda\cdot\vec J_{\gamma\eta}(0)|
\vec p_d\,1 m_d\rangle\,
\eeq
between the deuteron ground state with spin projection $m_d$ on the
$z$-axis and the final two-nucleon state
with total spin $s=0,1$, its projection $m_s$ and total
isospin $t=0,1$. The index $\lambda=\pm 1$ denotes the circular photon polarization.

The calculation of the matrix element (\ref{Tsms}) follows the standard
recipe in which the spectator model (IA) is treated
as the first basic approximation, and the interaction
between the final particles is taken into account in the form of
additional terms which are presented by the diagrams (b) through (d) in
Fig.~\ref{fig1}. Thus the resulting amplitude is the sum of the four contributions
\begin{equation}\label{fff}
T=T^{IA}+T^{NN}+T^{\eta N}+T^{\pi N}\,,
\end{equation}
where the last term corresponds to the pion exchange mechanism which
is the main object of the present investigation. The procedure, which
we used to calculate the first three terms, is described in a number
of papers~\cite{Sauermann,FiAr97,FiAr04,Sibir}, and we refer the reader
to these for more details.

The elementary $\eta$ photoproduction operator $t_{\gamma \eta}$,
entering the first three terms in Eq.~(\ref{fff}) is taken from the
isobar model EtaMAID \cite{EtaMAID}. The latter includes contributions
from Born terms, vector meson exchanges in the $t$-channel, and
$s$-channel resonances $D_{13}(1520)$, $S_{11}(1535)$, $S_{11}(1650)$,
$D_{15}(1675)$,  $F_{15}(1680)$, $D_{13}(1700)$, $P_{11}(1710)$, and
$P_{13}(1720)$. This model provides a reasonable description of the
available data on $\eta$ photo- and electroproduction on the nucleon
in the energy region up to a total c.m.\ energy $W=2$~GeV,
which corresponds to  a lab photon energy
$E_\gamma^{lab}=1650$~MeV. In the energy region of the
present study from threshold up to $E_\gamma^{lab}=800$~MeV, the
resonance $S_{11}(1535)$ dominates the cross
section, and the other resonances are of minor
importance only. The off-mass shell behavior of the photoproduction
operator was taken according to the energy-momentum conservation at
the single-nucleon vertex
\begin{eqnarray}
\vec{p}_{\,in}&=&\vec{q}_\eta+\vec{p}_1+\vec{p}_2-\vec{k}\,,\\
E_{\,in}&=&\omega_\eta+E_1+E_2-E_\gamma\,,\nonumber
\end{eqnarray}
where $(E_{\,in},\vec{p}_{\,in})$ denotes the four-momentum of the initial nucleon in the
deuteron. In all cases, the $\eta$ meson and both nucleons were taken on-shell.

The amplitude $T^{\pi N}$ in (\ref{fff}) has the form
\begin{equation}
T^{\pi N}=\langle s,\,m_s,\,t\,|\ t_{\pi \eta}(N_2)\,G_{\pi NN}\,t_{\gamma \pi}^\lambda (N_1)\,
|1 m_d \rangle -(-1)^{t+s}(1\leftrightarrow 2)\,,
\end{equation}
where $t_{\gamma \pi}^\lambda (N_i)$ and $t_{\pi\eta}(N_i)$, $i=1,2$ denote
the amplitudes for photoproduction of a pion  and for
$\pi N_i\to\eta N_i$ rescattering on a nucleon $N_i$, respectively.  $G_{\pi NN}$ is the
propagator of the three free particles $\pi$, $N_1$, and $N_2$. For
$t_{\gamma\pi}^\lambda $ we took the MAID2007 amplitudes
\cite{MAID2007}. With respect to its isospin
structure one has
\begin{equation}
t_{\gamma \pi}^\lambda =A^{(0)}-2A^{(-)}\,,\quad \mbox{for}\ \ t=0\,,
\end{equation}
and
\begin{equation}
t_{\gamma \pi}^\lambda =A^{(+)}+2A^{(0)}\,,\quad \mbox{for}\ \ t=1\,,
\end{equation}
with $t$ denoting the total isospin of the two nucleons. The amplitudes $A^{(0,\pm)}$
are those which appear in the isospin decomposition of the pion
photoproduction amplitude according to
\begin{equation}
%t(\gamma N\to \pi N)
t_{\gamma \pi}^\lambda (N)=A^{(+)}\delta_{b3}+\frac12 A^{(-)}[\tau_b,\tau_3]+A^{(0)}\tau_b\,,
\end{equation}
with $b$ being the isotopic index of the produced pion.

\begin{figure}
\begin{center}
\resizebox{0.8\textwidth}{!}{%
\includegraphics{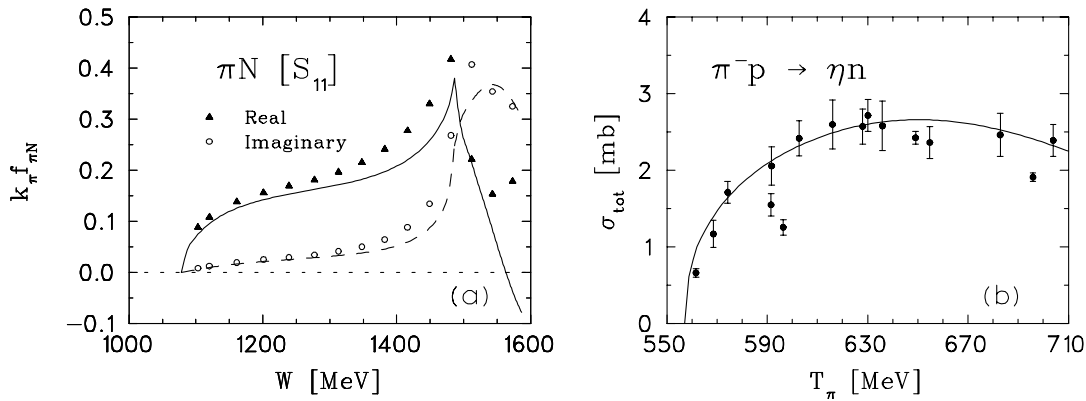}
} \caption{Left panel (a): $S_{11}$ partial wave amplitude f$_{\pi N}$ for $\pi N$
scattering predicted by our parametrization of the $S_{11}(1535)$ resonance [see
Eqs.~(\ref{35}) through (\ref{42})] as function of the total c.m.\ energy $W$. The
amplitude is multiplied with the pion momentum k$_\pi$. Notation:
Solid curve : real part, dashed: imaginary part. Circles and triangles
represent the VPI analysis \cite{VPI}. Right panel (b):
Total $\pi^-p\to\eta n$ cross section. The data are
taken from the compilation in Ref.~\cite{Zagreb}.} \label{fig2}
\end{center}
\end{figure}

In order to calculate $\eta N\to\eta N$ and $\pi N\to \eta N$
scatterings we use an isobar ansatz from Ref.~\cite{FiAr04}, which is
driven exclusively by the dominant $S_{11}(1535)$ resonance, parametrized in the form
\begin{equation}\label{35}
t_{\alpha\beta}(\omega_{\eta
  N},\vec{q},\vec{q}^{\,\prime})=\frac{g_\alpha(\vec{q}\,)\,g_\beta(\vec{q}^{\,\prime})}
{\omega_{\eta N}-M_0-\Sigma_\eta(\omega_{\eta
    N})-\Sigma_\pi(\omega_{\eta
    N})+\frac{i}{2}\Gamma_{\pi\pi}}\,,\quad \alpha,\beta\in
\{\pi,\eta\}\,,
\end{equation}
where the $S_{11}(1535)$ self energies,
$\Sigma_\eta$ and $\Sigma_\pi$, are given by
\begin{equation}
\Sigma_\alpha(\omega_{\eta
  N})=\frac{1}{2\pi^2}\int\frac{q^2dq}{2\omega_\alpha(q)}\frac{g^2_\alpha(q)}{\omega_{\eta
    N}-E_N(q)-\omega_\alpha(q)+i\epsilon}\,,
\quad \alpha\in\{\pi,\eta\}\,.
\end{equation}
Here, $\omega_{\eta N}$ denotes the invariant $\eta N$ energy, and $E_N(q)=\sqrt{q^2+M_N^2}$ and
$\omega_\alpha(q)=\sqrt{q^2+M_\alpha^2}$ the on-shell energies of
nucleon and meson $\alpha$, respectively. The vertex functions $g_\alpha(q)$ are
taken in a Hulth\'en form
\begin{equation}\label{42}
g_\alpha(q)=\frac{g_\alpha}{1+\big(q/\beta_\alpha\big)^2}\,,
\end{equation}
with $\beta_\alpha$ as a cut-off momentum.
The two-pion channel is included in a simplified manner by adding the
$S_{11}\to\pi\pi N$ decay width in the form
\begin{equation}
\Gamma_{\pi\pi}=\gamma_{\pi\pi}\frac{W-M_N-2m_\pi}{m_\pi}\,,
\end{equation}
with $\gamma_{\pi\pi}=4.3$~MeV, and
$M_N$ and $m_\pi$ denoting nucleon and pion masses, respectively.
The parameters $M_0$, $g_\alpha$, and $\beta_\alpha$ were adjusted in
such a way that the $\eta N$ scattering length
\begin{equation}\label{a}
a_{\eta N}=(0.5+i0.32)\ \mbox{fm}\,
\end{equation}
is reproduced. This may be considered on one side as an approximate average of
the scattering lengths given by modern $\eta N$ scattering
analyses~\cite{BeTa,Wycech,Zagreb}, and on the other side providing a reasonably
good description of the reactions in the channels coupled to the
$\eta N$ system as is shown in Fig.~\ref{fig2}. The fit parameters
are listed in Table 1 of Ref.~\cite{FiAr04}.

%%%%%%%%%%%%%%%%%%%%%%%%%%%%%%%%%%%%%%%%%%%%%%%%%%%%%%%%%%%%%%%%%
%%%%%%%%%%%%%%%%%%%%%%%%%%%%%%%%%%%%%%%%%%%%%%%%%%%%%%%%%%%%%%%%%

In order to evaluate the $NN$ scattering amplitude as well as the deuteron wave
function we have used the separable representation of the Paris
potential according to Ref.~\cite{Haiden}. All partial waves up to
$J=2$ were taken into account. Furthermore, in
order to check the  dependence of the results on the choice of the
$NN$ scattering model we carried out the same calculations with
the CD-Bonn model~\cite{CD} as well as with two versions of the
Nijmegen potential (Nijm93 and NijmII) from Ref.~\cite{Nijm} and found
the results to be very close to those obtained with the separable Paris
potential of Ref.~\cite{Haiden}. The deviation is less than 10~$\%$ in
the near threshold region and decreases rapidly with increasing
energy. This confirms the results of
Refs.~\cite{Sauermann,FiAr97,Sibir,FiAr04}  that the calculation is
practically insensitive to the choice of the $NN$ potential.

\section{Observables}\label{observables}
The general formalism for cross section and polarization observables
for photoproduction of a pseudoscalar meson on a deuteron
is presented in detail in Ref.~\cite{ArF05}. From this work we take
the general expressions for total and semi-inclusive differential
cross sections in the c.m.\ system, when only the direction of the
produced $\eta$ meson, characterized by
$\Omega_\eta=(\theta_\eta,\phi_\eta)$, is measured,
\beqa
\frac{d^2\sigma}{d\Omega_\eta}&=&
\frac{d^2\sigma_0}{d\Omega_\eta}
\Big[1
+P^d_1\,T^0_{11}\,\sin\phi_{\eta d}\,d^1_{10}(\theta_d)
+P^d_2\,\sum_{M= 0}^2 T^0_{2M}\,\cos M\phi_{\eta d}\,d^2_{M0}(\theta_d)
\nonumber\\&&
+P^\gamma_\ell\,\Big\{
\Sigma^\ell\,\cos 2\phi_\eta+
 P^d_1\,\sum_{M= -1}^1 T_{1M}^\ell\,\sin\psi_{M}\,d^1_{M0}(\theta_d)
+P^d_2\,\sum_{M= -2}^2 T_{2M}^\ell\,\cos\psi_{M}\,d^2_{M0}(\theta_d)
\Big\}
\nonumber\\&&
+P^\gamma_c\,\Big\{P^d_1\,\sum_{M= 0}^1 T_{1M}^c
\,\cos M\phi_{\eta d}\,d^1_{M0}(\theta_d)
+P^d_2\,\sum_{M= 1}^2 T^c_{2M}\,\sin M\phi_{\eta d}\,d^2_{M0}(\theta_d)
\Big\}\Big]\,.\label{diffcrossc}
\eeqa
Here $P^\gamma_\ell$ and $P^\gamma_c$ denote
the degrees of linear and circular photon polarization, and $P^d_1$ and
$P^d_2$ the deuteron vector and tensor polarization of the deuteron
target with respect to an axis characterized by
$\Omega_d=(\theta_d,\phi_d)$. Furthermore, we have abbreviated for
convenience $\phi_{\eta d}=\phi_{\eta}-\phi_{d}$ and
$\psi_{M}=M\phi_{\eta d}-2\phi_{\eta}$.
The unpolarized semi-inclusive differential cross section and the
associated asymmetries are given by
\beqa
\frac{d^2\sigma_0}{d\Omega_\eta}&=&
V_{00}^1(\theta_\eta)\,,\label{unpoldiff}\\
\Sigma^\ell\ell (\theta_\eta)\,\frac{d^2\sigma_0}{d\Omega_\eta}&=&
W_{00}(\theta_\eta)\,,\label{sigasy}\\
T_{11}^0(\theta_\eta)\,\frac{d^2\sigma_0}{d\Omega_\eta}&=&
-2\, Im\, [
V_{11}^1(\theta_\eta)]\,,
\label{tim}\\
T_{IM}^c(\theta_\eta)\,\frac{d^2\sigma_0}{d\Omega_\eta}&=&
-(2-\delta_{M0})\, Im \, [i^{-\delta_{I1}}\,
V_{IM}^1(\theta_\eta)]\,,\quad\mbox{for }0\leq M\leq I\,,
\label{timc}\\
T_{IM}^\ell (\theta_\eta)\,\frac{d^2\sigma_0}{d\Omega_\eta}&=&
i^{\delta_{I1}}\,W_{IM}(\theta_\eta)\,,
\quad\mbox{for }-I\leq M\leq
I\,.\label{timl}
\eeqa
The quantities $V_{IM}^1$ and $W_{IM}$ are hermitean quadratic forms
in the $T$-matrix elements integrated over the $\eta$ momentum
$q_\eta$ and the angle $\Omega_{p}$ of the relative momentum $\vec p
=(p,\Omega_{p})$ of the final two-nucleons according to
\beqa
W_{IM}(\theta_\eta)&=&
-\frac{\hat I}{\sqrt{3}}\,\int dq_\eta\,d\,\Omega_{p}\,c_{kin}\,
\sum_{m_d m_d'}(-)^{1-m_d}
\left(
\begin{matrix}
1&1&I \cr m_d'&-m_d&M \cr
\end{matrix} \right)\nonumber\\
&&
\sum_{s m_s t}t^*_{s m_s t 1 m_d'}(q_\eta,\, \theta_\eta,\, \theta_p,\, \phi_{p\eta})
\,t_{s m_s t -1 m_d}(q_\eta,\, \theta_\eta,\, \theta_p,\,
\phi_{p\eta})\,,\\
V_{IM}^1(\theta_\eta)&=&
\frac{\hat I}{\sqrt{3}}\,\int dq_\eta\,d\,\Omega_{p}\,c_{kin}\,
\sum_{m_d m_d'}(-)^{1-m_d}
\left(
\begin{matrix}
1&1&I \cr m_d'&-m_d&M \cr
\end{matrix} \right)\nonumber\\
&&
\sum_{s m_s t}t^*_{s m_s t 1 m_d'}(q_\eta,\, \theta_\eta,\, \theta_p,\, \phi_{p\eta})
\,t_{s m_s t 1 m_d}(q_\eta,\, \theta_\eta,\, \theta_p,\,
\phi_{p\eta})\,,
\eeqa
with $\phi_{p\eta}=\phi_{p}-\phi_{\eta}$.
Here the small $t$-matrix elements are defined by
\beq
t_{s m_s t \lambda m_d}=e^{-i(\lambda+m_d-m_s)\phi_\eta}\,T_{s m_s t \lambda
  m_d}\,,
\eeq
and $c_{kin}$ denotes a kinematic factor
\beq
c_{kin}=
\frac{1}{(2\pi)^5}\frac{E_d M_N^2p q^2_\eta}{4E_\gamma W \omega_{NN}\omega_\eta}\,,
\eeq
where the total energy in the $\gamma d$ c.m.\ system and the invariant
energy of the final two-nucleon subsystem are denoted by $W$ and
$\omega_{NN}$, respectively.

The total cross section with inclusion of photon and target
polarization effects is obtained by integrating
${d^2\sigma}/{d\Omega_\eta}$ over the meson spherical angle $\Omega_\eta$
and reads
\beqa
\sigma(P^\gamma_\ell,P^\gamma_c,P^d_1,P^d_2)
&=& \sigma_0\,\Big[1+P^d_2\,\overline T_{20}^{\,0}\,\frac{1}{2}
(3\cos^2\theta_d-1)
+P^\gamma_c\,P^d_1\,\overline T_{10}^{\,c}\,\cos\theta_d\nonumber\\&&
+P^\gamma_\ell\,P^d_2 \,\overline T_{22}^{\ell}\cos(2\phi_d)
\,\frac{\sqrt{6}}{4}\sin^2\theta_d\Big]\,,
\eeqa
where the unpolarized total cross section and the corresponding asymmetries
are given by
\beqa
\sigma_0&=&\int d\Omega_\eta
\,\frac{d^2\sigma_0}{d\Omega_\eta}\,,\\
\sigma_0\,\overline T_{IM}^{\,\alpha}&=&
\int d\Omega_\eta
\,\frac{d^2\sigma_0}{d\Omega_\eta}\,T_{IM}^{\,\alpha}\,,
\eeqa
with $\alpha\in\{0,\ell,c\}$.

\section{Results and discussion}\label{Results}

\subsection{Total cross section}

\begin{figure}[ht]
\begin{center}
\resizebox{1.\textwidth}{!}{%
\includegraphics{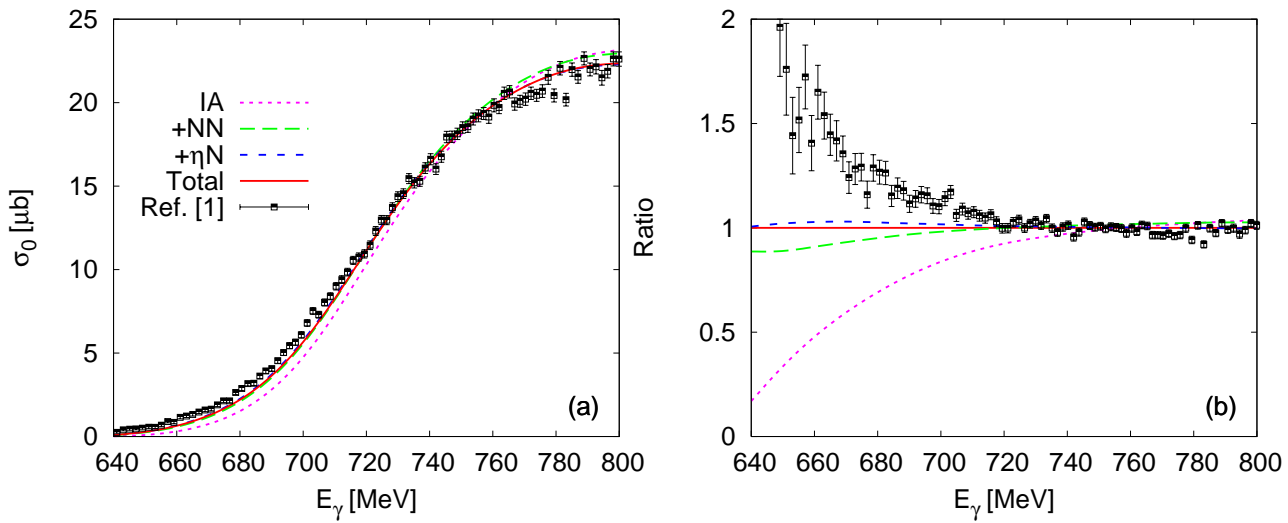}
} \caption{(Color online) Left panel (a): Unpolarized total cross section $\sigma_0$
  for $\gamma d\to\eta np$. The dotted (magenta), long-dashed (green)
  and short-dashed (blue) curves correspond to
  the impulse approximation (IA) and successive inclusion of $NN$ and
  $\eta N$ rescatterings, respectively. The solid curve (Total, red) includes also
  the pion exchange mechanism. Experimental data from Ref.~\cite{Krusche95}.
  Right panel (b): Ratios of the various approximations with respect to
  the ``Total'' one.} \label{fig3}
\end{center}
\end{figure}

We start the discussion with the total unpolarized cross section and
the associated asymmetries for polarized beam and target. The
theoretical results for the different approximations together with
available experimental data are presented in the left panel of
Fig.~\ref{fig3} whereas the right panel shows the ratios with respect
to the complete calculation.

In the near-threshold region
the interaction effects are very important as one can clearly see in
the %lower
panel (b). The resulting enhancement is about a factor 5 at the photon energy 640~MeV.
As was earlier noted in Refs.~\cite{FiAr97,FiAr04}, the reason of such
a strong influence of FSI is twofold. Firstly, the interaction between
the final particles allows to balance the strong mismatch
between the momentum needed for
$\eta$ production on a nucleon in the deuteron and the
characteristic internal nucleon momentum within the deuteron. We
recall that in the impulse approximation the $\eta$ meson is produced only through the
high momentum components of the target wave function, which appears in
the deuteron with a small probability. Thus, this momentum
mismatch is balanced mainly by the two-body $NN$ and to a lesser
extent by the $\eta N$ FSI, producing a very large
enhancement of the $\eta$ production rate. The second reason is
an appreciable attraction in the $NN$ as well as in the $\eta N$ system which leads to
an additional enhancement of the resulting cross section.

%%%%%%%%%%%%%%%%%%%%%%%%%%%%%%%%%%%%%%%%%%%%%%%%%%%%%%%%%%%%%%%%%
%%%%%%%%%%%%%%%%%%%%%%%%%%%%%%%%%%%%%%%%%%%%%%%%%%%%%%%%%%%%%%%%%

As for the pion exchange mechanism (diagram (d) of Fig.~\ref{fig1}),
it plays at lower energies only a minor role. Its inclusion
leads to a decrease of the cross section of about 3~$\%$ at $E_\gamma^{lab}=650$~MeV.
As already noted in the Introduction, this rather modest role of the pion
exchange is expected as a consequence of the large characteristic
momentum of the intermediate pion. Namely, if we assume that the largest
contributions come from states where the pion is approximately on-shell, than its
momentum is about 400~MeV/c. The corresponding mechanism is effective only at small
internucleon distances, about 0.5~fm, which are not essential for the incoherent channel. At the same time, as will be discussed in the next subsection,
these mechanisms should be rather effective in the region of
backward $\eta$ emission angles $\theta_\eta$ where large momentum transfers
naturally emphasize small internucleon distances.

As one readily notes in the right panel of Fig.~\ref{fig3}, even after inclusion of
all FSI effects our calculation systematically underestimates the data
in the region $E_\gamma^{lab}< 720$~MeV. This question was discussed in
quite some detail in Ref.~\cite{FiAr04}. In this work the reaction $\gamma
d\to \eta np$ was calculated within a three-body model, and it has
been shown that the perturbative approach, in which $\eta NN$ interaction
is reduced to pairwise $NN$ and $\eta N$ rescatterings, is unable to
provide an accurate description of the reaction dynamics at low
energy. The main reason for this fact is a relatively strong attraction
which generates virtual poles in the $s$-wave $\eta NN$ states
$J^\pi=1^-,\ T=0$ and $J^\pi=0^-,\ T=1$~\cite{FiAretaNN}. Obviously,
the corresponding singularities in the $\eta NN$ scattering amplitude
cannot be generated by a truncated perturbation expansion. According
to the results of Refs.~\cite{FiAr00,FiAr04}, the  inclusion of
"three-body" effects leads to a visible improvement of the theory and
brings the calculation into a better agreement with the
data. Nevertheless, even in this case the theoretical total cross
section still underestimates the data of
Ref.~\cite{Krusche95} slightly.

At higher energies the effect of pion exchange becomes comparable in
size to that coming from $\eta N$ rescattering. Despite the
already mentioned rather large characteristic momentum of the
intermediate pion, which reduces its contribution near threshold, the relatively
large value of the pion photoproduction cross section at these higher
energies appears to compensate this effect and makes the pion exchange
more important although still quite small in the unpolarized total
cross section.

%%%%%%%%%%%%%%%%%%%%%%%%%%%%%%%%%%%%%%%%%%%%%%%%%%%%%%%%%%%%%%%%%
\begin{figure}
\begin{center}
\resizebox{1.\textwidth}{!}{%
\includegraphics{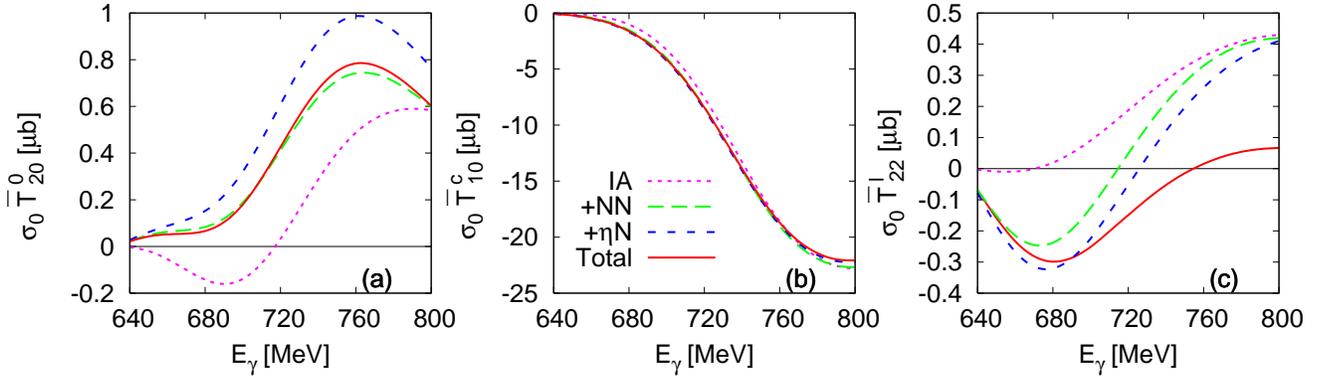}
} \caption{(Color online) Polarization asymmetries of the   total cross section for
  $\gamma d\to\eta np$: (a)   $\sigma_0\,\overline T_{20}^{\,0}$, (b)
  $\sigma_0\,\overline  T_{10}^{\,c}$, and (c) $\sigma_0\,\overline
  T_{22}^{\,\ell}$. Notation as in Fig.~\ref{fig3}.} \label{fig3a}
\end{center}
\end{figure}
%%%%%%%%%%%%%%%%%%%%%%%%%%%%%%%%%%%%%%%%%%%%%%%%%%%%%%%%%%%%%%%%%

Fig.~\ref{fig3a} shows the influence of the FSI effects on the
integrated polarization observables $\sigma_0\overline T^c_{10}$,
$\sigma_0\overline T^0_{20}$, and $\sigma_0\overline T^l_{22}$. As
mentioned above, these are the only polarization observables
contributing to the total cross section. The largest one is
$\sigma_0\overline T^c_{10}$ for circular photon
polarization and a vector polarized deuteron target shown in panel (b)
of Fig.~\ref{fig3a} . However, the influence of
FSI on $\sigma_0\overline T^c_{10}$ is rather weak and comparable in
size to that noted for the total cross sections. This observable
determines the contribution of eta photoproduction to the
Gerasimov-Drell-Hearn sum rule (GDH)~\cite{GDH}.

For the other two observables in panels (a) and (c) of
Fig.~\ref{fig3a} the FSI influence is much more notable. For
$\sigma_0\overline T^0_{20}$ (panel (a) of Fig.~\ref{fig3a}) for
unpolarized photons and a tensor
polarized target the contributions from $\eta N$ rescattering and pion
exchange appear to compensate each other, leaving $NN$ rescattering as
the main effect. Finally, $\sigma_0\overline T^\ell_{22}$ needing
linearly polarized photons and again a tensor polarized target, $NN$
rescattering is dominant at low energies but dies out rapidly with
increasing energy whereas the pion exchange contribution shows just
the opposite behavior becoming dominant at higher energies. The
contribution from $\eta N$ rescattering remains moderate in the whole
energy range. Thus, these latter two observables would provide an
interesting test for the study of the various FSI effects.

\subsection{Semi-inclusive differential cross section}

%%%%%%%%%%%%%%%%%%%%%%%%%%%%%%%%%%%%%%%%%%%%%%%%%%%%%%%%%%%%%%%%%
\begin{figure}
\begin{center}
\resizebox{1.\textwidth}{!}{%
\includegraphics{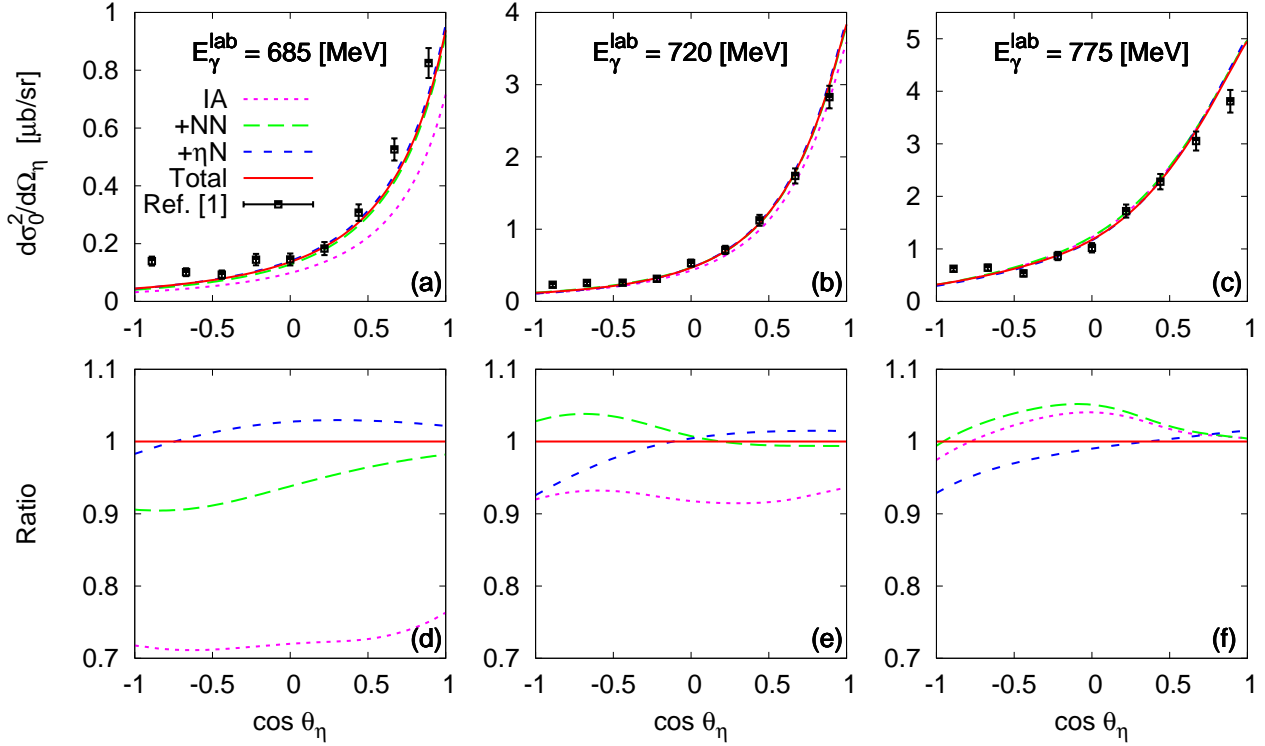}}
\caption{(Color online) Semi-inclusive unpolarized differential cross section
  for $\gamma d\to\eta np$ at the energies $E_\gamma^{lab}= 685$~MeV (a),
  720~MeV (b) and 775~MeV (c). The data are from Ref.~\cite{Krusche95}. The lower
  panels show the corresponding ratios of the cross sections with
  respect to the complete calculation ``Total''. Notation as in
Fig.~\ref{fig3}.} \label{fig4}
\end{center}
\end{figure}
%%%%%%%%%%%%%%%%%%%%%%%%%%%%%%%%%%%%%%%%%%%%%%%%%%%%%%%%%%%%%%%%%
As next we will consider the semi-inclusive differential cross section
of Eq.~(\ref{diffcrossc}),
for which only the direction of the produced $\eta$ meson is detected.
In Fig.~\ref{fig4} the angular distributions predicted by our
calculation are plotted for three energies together with the experimental
data of Krusche et al.~\cite{Krusche95}.
As expected the FSI effects are most notable in the region of backward angles, where
the momentum transfer is maximal and, as a result, the two-nucleon response
becomes important. To show the relative size of the different FSI
mechanisms we plot in the lower panels of Fig.~\ref{fig4} the ratios
of the various approximations with respect to the complete
calculation, labelled ``Total''. One readily notes that at backward
angles $\theta_\eta$ the inclusion of FSI effects leads to a sizable
increase over the IA cross section. For energies $E_\gamma \geq
720$~MeV and at backward angles, i.e.\ in the region
$\cos\theta_\eta\leq -0.5$, the effect of pion exchange becomes
comparable in size to $\eta N$ rescatterings, but acting in the
opposite direction. On the other hand, the angular
distributions clearly show that the $\eta$ meson is produced mostly
at forward direction, where the influence of all FSI mechanisms
remains small and thus the IA works
quite well (except in the region below 720~MeV).

\begin{figure}[h]
\begin{center}
\resizebox{.9\textwidth}{!}{%
\includegraphics{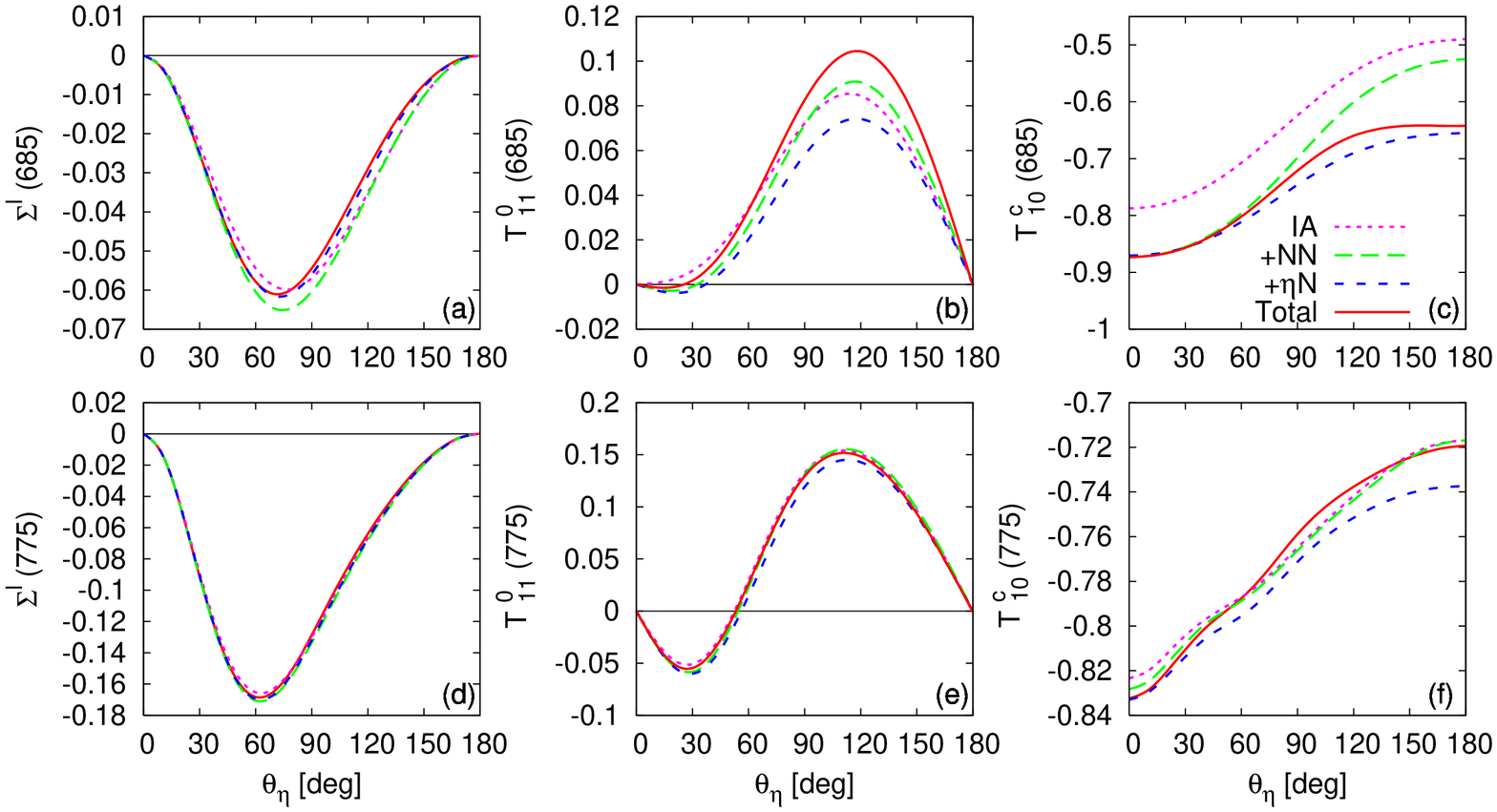}}
\caption{(Color online) Polarization observables $\Sigma^\ell $, $T^0_{11}$,
  and   $T^c_{10}$ of the
  semi-inclusive differential cross section for $\gamma d\to\eta
  np$ at $E_\gamma^{lab}=685$ (panels (a) through (c)) and 775~MeV
  (panels (d) through (f)). Notation as in
  Fig.~\ref{fig3}. } \label{fig4a}
\end{center}
\end{figure}

\begin{figure}
\begin{center}
\resizebox{.9\textwidth}{!}{%
\includegraphics{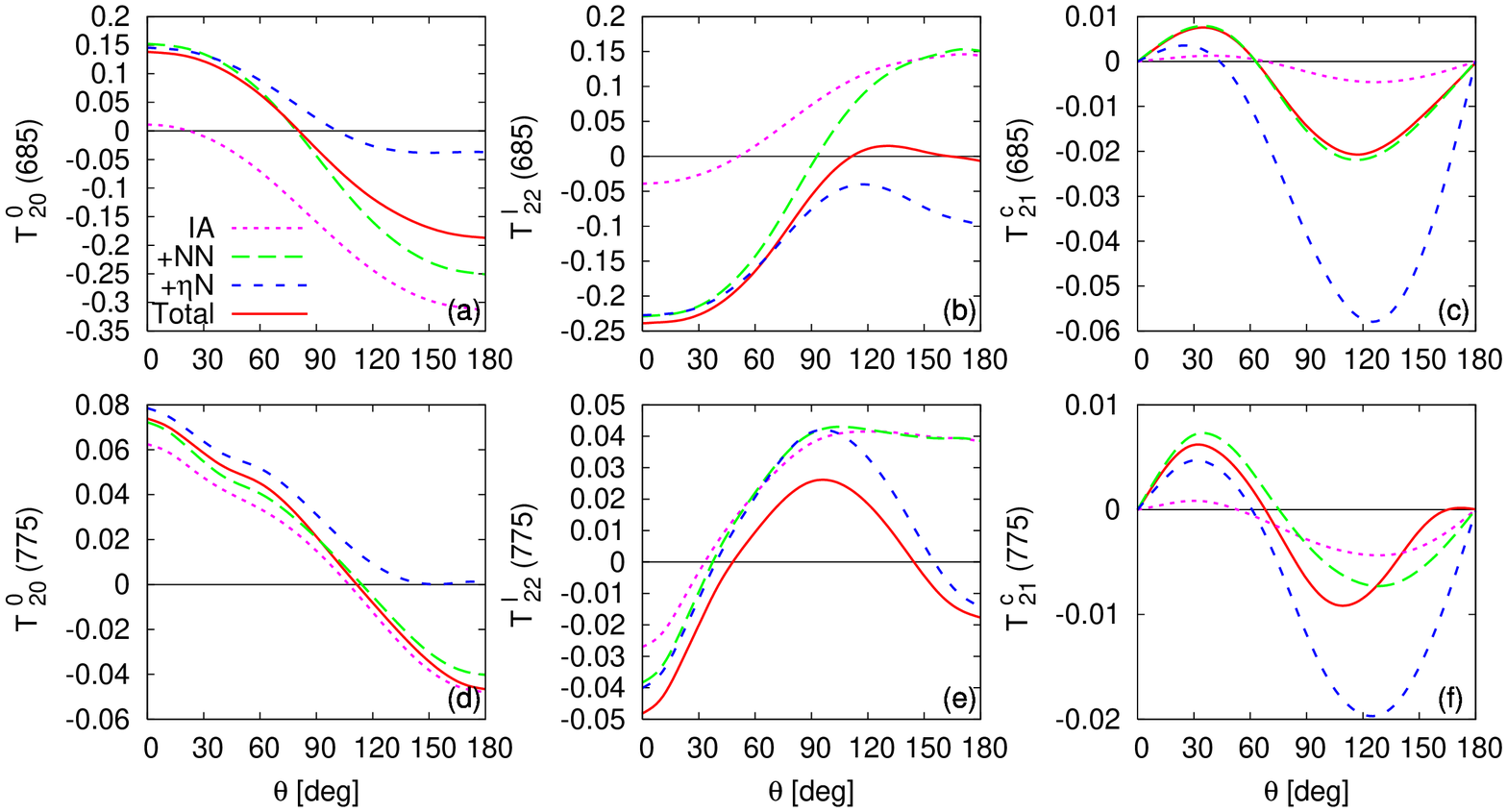}
} \caption{(Color online) Polarization observables $T^0_{20}$,
  $T^\ell_{22}$ and $T^c_{21}$ of the
  semi-inclusive differential cross section for $\gamma d\to\eta
  np$ at $E_\gamma=685$ and 775~MeV. Notation as in Fig.~\ref{fig3}. } \label{fig4b}
\end{center}
\end{figure}

In Figs.~\ref{fig4a} and \ref{fig4b} we present for two energies a few
polarization observables of the differential cross section in
Eq.~(\ref{diffcrossc}), which are in general more sensitive to
dynamical details of the reaction. We have chosen those asymmetries
which have relatively large values in the energy region, considered.

%%%%%%%%%%%%%%%%%%%%%%%%%%%%%%%%%%%%%%%%%%%%%%%%%%%%%%%%%%%%%%%%%
%%%%%%%%%%%%%%%%%%%%%%%%%%%%%%%%%%%%%%%%%%%%%%%%%%%%%%%%%%%%%%%%%

Fig.~\ref{fig4a} exhibit $\Sigma$, the asymmetry for linearly
polarized photons and an unpolarized target, $T^0_{11}$ for
unpolarized radiation and a vector polarized deuteron target, and
$T^c_{10}$, the asymmetry for circular photon polarization and again a
vector polarized target. The latter asymmetry determines the GDH sum rule,
mentioned above. The linear photon asymmetry $\Sigma$ is very little
affected by FSI already at the lower energy of 685~MeV (panel (a)) and
even less at 775~MeV (panel (d)). The other two observables in
Fig.~\ref{fig4a} are more sensitive to FSI, at least for the lower
energy.  For $T^0_{11}$ the influence of $NN$
rescattering at 685~MeV (panel (b)) is quite small in contrast to a
significant decrease of
the IA by $\eta N$ rescattering. This decrease is more than
counterbalanced by the pion exchange having the largest contribution
and which leads to a significant overall increase compared to the
IA. At the higher energy this sensitivity is lost (panel (e)). Turning
to $T^c_{10}$ in panel (c) one readily notes that for this lower energy
$NN$ rescattering is dominant at small angles, where the other effects
are negligible, leading to an overall negative increase. At larger
angles $\eta N$ rescattering becomes the dominant effect yielding a
further sizeable negative increase at 180$^\circ$. Pion exchange is
small and acts opposite to $\eta N$ rescattering. At the higher energy all
interaction effects become considerably smaller (see panel (f)). Pion
exchange becomes comparable to $\eta N$ rescattering but almost
cancels the latter, so that the overall FSI influence is very small.

 The polarization observables shown in Fig.~\ref{fig4b} for a tensor
polarized deuteron target exhibit a stronger sensitivity to interaction
effects, in particular for the lower energy (panels (a) through
(c)). For the two observables  $T^0_{20}$ and $T^\ell_{22}$ in panels
(a) and (b) $NN$ rescattering is again dominant at forward angles with
all other contributions negligible. However, at backward angles both
$\eta N$ rescattering and pion exchange become quite sizeable, but
cancelling each other partially. This cancellation is complete for
$T^c_{21}$ in panel (c), leaving as net effect $NN$ rescattering. At
the higher energy of 775~MeV the overall influence of FSI diminishes
substantially for $T^0_{20}$ in panel (d), while they remain stronger
for $T^\ell_{22}$ and $T^c_{21}$ (panels (e) and (f)). In particular
$\eta N$ rescattering appears quite strong for backward angles in both
observables. But also pion exchange is sizeable.

%%%%%%%%%%%%%%%%%%%%%%%%%%%%%%%%%%%%%%%%%%%%%%%%%%%%%%%%%%%%%%%%%
%%%%%%%%%%%%%%%%%%%%%%%%%%%%%%%%%%%%%%%%%%%%%%%%%%%%%%%%%%%%%%%%%

 In general, these polarization observables exhibit quite an
interesting sensitivity to interaction effects, in many cases quite
stronger than for the unpolarized angular distribution and thus offer
an interesting possibility to study this reaction in much greater detail.

\section{Conclusion}\label{Conclusion}

In the present work we have considered incoherent photoproduction of
$\eta$ mesons on the deuteron in the energy region from threshold up
to $E_\gamma^{lab}=800$~MeV, with inclusion of the most important FSI
mechanisms, $NN$ and $\eta N$ rescattering, as well as the pion
exchange contribution ($\pi N$). Our main purpose was to
determine the role of the pion exchange without resorting to the
approximations used in previous work. In general, as has also
been shown by previous studies, the role of FSI in this reaction
is rather unimportant for the total unpolarized cross section, except
in the region just above threshold  ($E_\gamma^{lab}<720$~MeV). But
this conclusion is not valid for the associated tensor target
asymmetries, where FSI effects become significant, i.e.\ for
$\overline T_{20}^{\,0}$ and $\overline T_{22}^{\,\ell}$, 
the latter for linearly polarized photons. 

Furthermore, sizeable FSI effects become significantly visible also at
higher energies in the differential cross section in those kinematical
regions, where the spectator-nucleon mechanism is suppressed, for
example for backward eta emission angles in the semi-inclusive
differential cross section. 

We also considered the role of FSI effects on different
polarization observables. As our calculation shows, some of these
observables, in particular $T^c_{10}$, $T^0_{20}$, and $T^l_{22}$
appear to be relatively large in the energy region considered in this
work and, at the same time, are rather sensitive to FSI effects.

As a general conclusion, we would like to stress the fact that in the region
$E_\gamma^{lab}\geq 720$~MeV the pion exchange mechanism appears to be of
the same size as $NN$ and $\eta N$ rescattering and should be taken
into account using a reliable model. Furthermore, it remains to be
seen in the futur whether the inclusion of pion degrees of freedom within a
more realistic three-body approach than in Ref.~\cite{FiArNuPh},
improves the agreement between theory and experiment in the
near-threshold region. 

\section*{Acknowledgment}
The work was supported by the Deutsche Forschungsgemeinschaft (SFB 1044). A.~Fix also
acknowledges support from the Dynasty foundation, the TPU grant
LRU-FTI-123-2014 and the MSU program 'Nauka' (project 3.825.2014/K).
M.~Tammam would like to thank the Egyptian Government for supporting
him through the Post Doctor Program of Egypt. A.~Fix and M.~Tammam
are also grateful for the kind hospitality of the Institut f\"ur Kernphysik of
the  Johannes Gutenberg-Universit\"at Mainz.

\end{document}